\documentclass[twocolumn,showpacs,pra,amsmath,amssymb]{revtex4}
\usepackage{graphicx}
\usepackage{bm}

\newcommand{\ket}[1]{|#1\rangle} 
\newcommand{\bra}[1]{\langle #1|}

\begin{document}

\title{Robustness of the adiabatic quantum search}
\author{Johan {\AA}berg}
\email{johan.aaberg@kvac.uu.se}
\author{David Kult}
\email{david.kult@kvac.uu.se}
\author{Erik Sj\"oqvist}
\email{erik.sjoqvist@kvac.uu.se}
\affiliation{Department of Quantum Chemistry,
Uppsala University, Box 518, SE-751 20 Uppsala, Sweden}

\begin{abstract}
The robustness of the local adiabatic quantum search to decoherence in the
instantaneous eigenbasis of the search Hamiltonian is examined. We
demonstrate that the asymptotic time-complexity of the ideal closed
case is preserved, as long as the Hamiltonian dynamics is present. In
the special case of pure decoherence where the environment monitors
the search Hamiltonian, it is shown that the local adiabatic quantum
search performs as the classical search.
\end{abstract}

\maketitle

Although the adiabatic approach to quantum computation
\cite{farhi00,farhi01} seems to differ significantly from the 
traditional circuit model, it has been proved that these two models
are, in a certain sense, equivalent \cite{aharonov04}. However, this
equivalence does not concern the robustness to noise, relaxation, or
decoherence.  Since the adiabatic schemes operate close to the energy
ground state it seems natural to guess that the adiabatic quantum
computer should be robust against relaxation effects
\cite{farhi01}. The alleged resistance to noise has been examined by
analytic means in Ref.~\cite{roland04} and it has been argued that
adiabatic quantum computers should be robust to decoherence
\cite{farhi01,kaminsky02}. Unitary control errors and resistance to
decoherence have been numerically investigated in
Ref.~\cite{childs01}.

In this paper, we examine the local adiabatic search algorithm
\cite{roland02,vanDam01} in the presence of decoherence in the instantaneous
energy eigenbasis \cite{paz99}. We demonstrate analytically a
robustness to this particular form of decoherence in the sense that
the asymptotic time-complexity of the ideal closed case is preserved,
no matter how small the Hamiltonian contribution is to the
dynamics. Only in the wide-open case
\cite{percival94}, where the Hamiltonian part is completely absent,
there is a difference in the time-complexity. 

Adiabatic quantum computation works by keeping the system close to the
ground state of a time-dependent Hamiltonian. This feature is in
contrast with, e.g., holonomic implementations of quantum gates
\cite{zanardi99}, which share the feature of adiabatic evolution, but
where it is essential that the gate can operate on arbitrary
superpositions without too large errors. For the functioning of the
adiabatic quantum computer in the presence of decoherence, on the
other hand, it is sufficient to require that the probability of
finding the system in the instantaneous ground state of $H(s)$ is
conserved.
This can be seen as one possible
generalization of the concept of adiabaticity to open systems. In this
generalized sense the wide-open case has an adiabatic limit, although
the Hamiltonian dynamics is absent. One may note that the wide-open
case can be seen as a quantum computational scheme in its own right, a
``wide-open adiabatic quantum computer'', where the dynamics is
governed by pure decoherence. A different approach to the concept of 
adiabaticity for open system has been put forward in Ref.
\cite{Sarandy}, and applied to adiabatic quantum computing in Ref. 
\cite{Sarandy2}.

The $N$-element search problem consists of finding a single marked
element in a disordered $N$-element list. The search problem is
associated with an $N$ dimensional Hilbert space with orthonormal basis 
$\{\ket{k} \}_{k=1}^N$, where the marked item corresponds to 
$\ket{\mu}\in\{ \ket{k} \}_{k=1}^N$. 
Following Refs.~\cite{farhi00,roland02,vanDam01},
we consider the family of Hamiltonians
\begin{equation}
\label{family} 
H(s) = -(1-s)\ket{\psi} \bra{\psi} - s \ket{\mu} \bra{\mu},
\end{equation}
where
\begin{equation}
\label{eqalsuperpos}
\ket{\psi} = \frac{1}{\sqrt{N}}\sum_{k=1}^N \ket{k} 
\end{equation}
and $s=t/T \in [0,1]$, $T$ being the run-time of the search. If the
evolution is adiabatic and we start in the energy ground state
$\ket{\psi}$, this family of Hamiltonians takes us to the marked state
$\ket{\mu}$ and thus solves the search problem. The only relevant
subspace is spanned by $\ket{\psi}$ and $\ket{\mu}$.  We denote the
instantaneous eigenvalues and orthonormal eigenvectors of $H(s)$
restricted to the relevant subspace by $E_n(s)$ and $\ket{E_n(s)}$,
respectively, where $n=0,1$. We further define
\begin{eqnarray} 
\Delta (s) & = & E_1 (s) - E_0 (s) = 
\sqrt{\frac{1+(N-1)(2s-1)^2}{N}}  
\end{eqnarray}
and
\begin{eqnarray} 
Z(s) & = & \big| \langle \dot{E}_0 (s) | E_1 (s) \rangle \big| =
\frac{\sqrt{N-1}}{1+(N-1)(2s-1)^2} .  
\end{eqnarray}
A useful property of $Z$ is 
\begin{equation}
\label{Zegensk}
\int_{0}^{1}Z(s)ds \leq\frac{\pi}{2},
\end{equation}
for all $N$. 

Decoherence in the instantaneous energy eigenbasis is modeled by the
master equation
\begin{eqnarray}
\label{eq:master}
\frac{d}{ds} \rho (s) & = & 
-iAT[H(s),\rho (s)] \nonumber \\
& & -BT[W(s),[W(s),\rho (s)]], 
\end{eqnarray}
where $A\geq 0$ and $B \geq 0$ are constants independent of $N$.
Here, $W(s)$ is assumed to be Hermitian, nondegenerate, and fulfill
$[W(s),H(s)]=0$. Furthermore, let $w_n(s)$ be the eigenvalues of
$W(s)$ corresponding to the eigenvectors $|E_n(s) \rangle$ and define
$\Gamma (s) = w_1 (s) - w_0 (s)$.

Next, we implement the idea of local adiabatic search \cite{roland02}
by making a monotone, sufficiently smooth reparametrization
$s\in [0,1] \rightarrow r=f(s) \in [0,1]$ of $H(s)$ and $W(s)$ in 
such a way that more time is spent near the minimum energy gap. 
In the closed case ($B=0$), it was shown in Ref. \cite{roland02} 
that the optimal choice    
\begin{eqnarray}
\label{Ldef}
f^{-1}(r) & = & 
\frac{1}{L} \int_{0}^{r}\frac{1}{\Delta^{2}(r')}dr', 
\nonumber \\
L &= & \int_{0}^{1}\frac{1}{\Delta^{2}(r')}dr' = 
\frac{N}{\sqrt{N-1}}\arctan(\sqrt{N-1}) 
\nonumber \\ 
 & \leq & \frac{\pi}{2}\frac{N}{\sqrt{N-1}} 
\end{eqnarray}
yields the criterion $T\gg \sqrt{N}$ for the run-time, in analogy with
the Grover search \cite{grover97}. Applying this reparametrization results
in the transformation $s\rightarrow r$ as well as in multiplication by
$\frac{df^{-1}}{dr}(r)$ of the right-hand side of
Eq.~(\ref{eq:master}).  Assume that $\rho (0) = \ket{E_0(0)}
\bra{E_0(0)}$ and let 
\begin{eqnarray}
Y(r) &=& \bra{E_0(r)} \rho(r) \ket{E_0(r)} - \bra{E_1(r)} \rho(r)
\ket{E_1(r)} \nonumber \\ &\equiv & \rho_{00}(r)-\rho_{11}(r).
\end{eqnarray}

We now address the main objective of this paper, which is to determine
how the probability to remain in the ground state depends on the
run-time $T$ and the parameters $A$ and $B$. The strategy is to
express this probability, indirectly in terms of $Y(r)$, as an
integral equation. Thereafter, we apply appropriate estimates to
obtain a lower bound for the probability.

We may rewrite Eq. (\ref{eq:master}) as an integral equation 
that takes the form 
\begin{equation}
\label{diff}
1-Y(r) =  4I(r),
\end{equation}
where 
\begin{eqnarray}
\label{gIdef} 
I(r) & = & \frac{1}{2}I_{+}(r) + \frac{1}{2}I_{-}(r), 
\nonumber \\
I_{\pm}(r) & = & 
\int_{0}^{r} e^{-T[BQ(r')\pm iAR(r')]}Z(r')u_{\pm}(r')dr', 
\nonumber \\
u_{\pm}(r') & = & 
\int_{0}^{r'}e^{T[BQ(r'')\pm iAR(r'')]}Z(r'')Y(r'')dr'',
\end{eqnarray}
and 
\begin{eqnarray}
\label{eq:qrdef}
Q(r) & = & \int_{0}^{r} \Gamma^{2}(r') \frac{df^{-1}(r')}{dr'} dr' = 
\frac{1}{L}\int_{0}^{r}\frac{\Gamma^{2}(r')}{\Delta^{2}(r')} dr', \nonumber 
 \\ R(r) & = & \int_{0}^{r} \Delta(r') \frac{df^{-1}(r')}{dr'} dr' =
\frac{1}{L}\int_{0}^{r}\frac{1}{\Delta(r')}dr' . 
\end{eqnarray}
We further define 
\begin{equation}
\label{zetadef}
\zeta = \min_{r\in [0,1]}\frac{\Gamma^{2}(r)}{\Delta(r)}.
\end{equation}

In the case where $A>0$, we wish to estimate $\left|1-Y(r)
\right|$. This can be done by calculating an upper bound for
$|I_{\pm}(r)|$, using $Z(r)\Delta(r) \leq \sqrt{(N-1)/N}$,
$\exp[-TBQ(r)]\leq 1$, $\exp\{-TB[Q(r)-Q(r')]\}\leq 1$ if $r\geq r'$,
as well as Eqs. (\ref{Zegensk}) and (\ref{zetadef}), which result in
\begin{eqnarray}
\label{tiIabs}
\left|I_{\pm}(r)\right| & \leq & \frac{\pi
L}{T}\sqrt{\frac{N-1}{N}}\frac{1}{\sqrt{B^{2}\zeta^{2}+ A^{2}}}\\ & &
+\frac{L}{T}\frac{\pi}{2}\int_{0}^{r}\left|\frac{d}{dr'}\left(\frac{Z(r') 
\Delta(r')}{B\frac{\Gamma^{2}(r')}{\Delta(r')}\pm iA}\right)\right| dr'.
\nonumber 
\end{eqnarray}
By use of $Z(r)\Delta^{2}(r) = \sqrt{N-1}/N$ and Eq.~(\ref{zetadef}),
the integral on the right-hand side of Eq.~(\ref{tiIabs}) can be
estimated as
\begin{eqnarray}
\label{absderiv}
 & & \int_{0}^{r}\left| \frac{d}{dr'} 
\left( \frac{Z(r') \Delta(r')}{B\frac{\Gamma^{2}(r')}{\Delta(r')} 
\pm iA} \right) \right| dr' 
\nonumber\\ 
 & \leq & \frac{A}{B^{2}\zeta^{2} + A^{2}}\int_{0}^{1}Z(r') 
\left|\frac{d}{dr'}\Delta(r')\right| dr'  
\nonumber\\ 
 & & + \frac{B}{B^{2}\zeta^{2} + A^{2}} 
\int_{0}^{1}Z(r')\left|\frac{d}{dr'}\Gamma^{2}(r')\right| dr' . 
\end{eqnarray}
Note that we have extended the integration
interval from $[0,r]$ to $[0,1]$. Since both $Z(r)$ and $\Delta(r)$
are symmetric around $r = \frac{1}{2}$, it follows that
$Z(r)|\frac{d}{dr}
\Delta(r)|$ has the same symmetry. Moreover, $\Delta(r)$ is increasing 
on the interval $[\frac{1}{2},1]$. Hence, $Z(r)|\frac{d}{dr}
\Delta(r)| = Z(r) \frac{d}{dr} \Delta (r)$ on $[\frac{1}{2},1]$, which
leads to
\begin{eqnarray}
\label{ZderD}
\int_{0}^{1}Z(r)\left|\frac{d}{dr} \Delta (r)\right|dr& =&  2\int_{1/2}^{1}Z(r)
\frac{d}{dr}\Delta (r)dr \nonumber \\ 
&\leq&  2\sqrt{\frac{N-1}{N}} \leq 2.
\end{eqnarray}
To deal with the second term on the right-hand side of Eq. 
(\ref{absderiv}), we introduce the following condition 
\begin{equation}
\label{condition}
\int_{0}^{1}Z(r)\left|\frac{d}{dr}\Gamma^{2}(r)\right| dr \leq K,
\end{equation}
where $K$ is a constant independent of $N$ \cite{thecondition}. Since
$B\Gamma^2(r)$ can be seen as the instantaneous strength of the
decoherence, the condition in Eq.~(\ref{condition}) essentially states
that the fluctuations in strength are not allowed to grow with $N$.
If one assumes that $\Gamma(r)=\eta\biglb(\Delta(r)\bigrb)$, where
$\eta : (0,\infty)
\rightarrow (0,\infty)$ is an increasing, sufficiently smooth function, 
it can be shown that it is sufficient that $\eta(x)\leq Cx^{\sigma}$,
where $C$ and $\sigma \geq \frac{1}{2}$ are constants, to fulfill the
condition in Eq.~(\ref{condition}). This means that the condition is
fulfilled for the particular case where $W(r)=H(r)$. By combining
Eqs. (\ref{Ldef}), (\ref{tiIabs}) - (\ref{condition}), and using that
$B^{2}\zeta^{2} + A^{2} \geq A^{2}$ and $B^{2}\Gamma^{4}(0)+ A^{2}
\geq A^{2}$, we obtain
\begin{eqnarray}
\label{soloc}
\rho_{00}(r) & \geq & 
1-2\pi^{2}\frac{ \sqrt{N}}{T} \left( \frac{1}{A} +
\sqrt{\frac{N}{N-1}}\frac{KB}{A^{2}} \right) .
\end{eqnarray}
Hence, it is a sufficient condition for local adiabaticity that $T\gg
\sqrt{N}$. In conclusion, an increased degree of eigenbasis
decoherence does not change the asymptotic behavior of the run-time of
the adiabatic search. This result is independent of the explicit form
of $W(r)$ as long as Eq. (\ref{condition}) is fulfilled and $[W(r),H(r)]=0$.

In the wide-open case \cite{percival94} ($A=0$), the protective effect
of the Hamiltonian dynamics is absent and one may expect that the
asymptotic behavior depends on the explicit choice of $W(r)$. To
verify this point, we let $W(r)$ be such that $\Gamma(r) =
\Delta^{\sigma}(r)$, $\sigma \geq 1$ \cite{sigmaremark}. We further
put $B=1$ for convenience. Notice that the choice $W(r) = H(r)$
corresponds to $\Gamma(r) = \Delta(r)$. We prove that in the wide-open
case with $\Gamma(r)=\Delta^{\sigma}(r)$, a sufficient and necessary
condition for adiabaticity is $T\gg N^{\sigma}$.  Note that we have to
show that the sufficient condition is also necessary, as we wish
to prove that the wide-open case is essentially different from the
$A\neq 0$ case.

To prove the sufficiency, we insert $\Gamma(r)=\Delta^{\sigma}(r)$ 
into Eq. (\ref{eq:qrdef}), and use that $\Delta(r)\geq 1/\sqrt{N}$ 
and $\sigma \geq 1$, to obtain 
\begin{eqnarray}
\label{wQWH}
Q(r) = \frac{1}{L} \int_{0}^{r} \Delta^{2\sigma-2}(r')dr' \geq 
\frac{1}{LN^{\sigma-1}}r 
\end{eqnarray}
Inserting Eq.~(\ref{wQWH}) into 
Eq.~(\ref{gIdef}) gives 
\begin{eqnarray}
\label{olikhet}
1-Y(r) & \leq & 4 \int_{0}^{r} \int_{0}^{r'} 
e^{-\frac{T}{LN^{\sigma-1}}(r'-r'')}Z(r')Z(r'')dr''dr'. 
\nonumber \\ 
\end{eqnarray}
Finally, we use $Z(r)\leq \sqrt{N-1}$ and Eq.~(\ref{Zegensk}) to obtain    
\begin{equation}
\label{wioloc}
\rho_{00}(r) \geq 1 -\frac{\pi^2}{2}\frac{N^{\sigma}}{T}.
\end{equation}
Thus, a sufficient condition for adiabaticity is $T \gg  N^{\sigma}$.
 
Now we show that this condition is also necessary. We let $Y_{T}$
denote the solution of Eq.~(\ref{diff}) for a given run-time $T$.
It can be proved that \cite{aberg04}
\begin{eqnarray}
\label{olikheten}
Y_{T}(r) \geq Y_0(r), 
\end{eqnarray}
which means that an evolution with non-zero run-time remains closer to
the instantaneous ground state than the evolution with zero
run-time. Insert Eq.~(\ref{olikheten}) into Eq.~(\ref{gIdef}) and
combine with Eq.~(\ref{diff}) to obtain
\begin{equation}
\label{grdlolik}
1- Y_{T}(1) \geq I_{0} ,
\end{equation}
where
\begin{eqnarray}
\label{Iintegdef}
I_{0} = 4 \int_{0}^{1}\int_{0}^{r} e^{-T[Q(r)-Q(r')]}Z(r) Z(r')
Y_{0}(r') dr' dr
\nonumber \\ 
\end{eqnarray}
and 
\begin{eqnarray}
\label{Ynolikhet}
Y_{0}(r) & = & \frac{1-(N-1)(2r-1)}{\sqrt{N}\sqrt{1 + (N-1)(2r-1)^{2}}}
\nonumber \\ 
 & \geq & -\frac{(N-1)(2r-1)}{\sqrt{N}\sqrt{1 + (N-1)(2r-1)^{2}}} .
\end{eqnarray}
In order for $Y_{T}(1) \rightarrow 1$, $I_0$ has to go to zero, 
since $Y_T(r) \leq 1$. Hence, we have found a necessary condition for the 
system to approach adiabaticity.

In order to express this necessary condition in terms of the run-time
$T$, let us use Eq.~(\ref{Ynolikhet}) in Eq.~(\ref{Iintegdef}) and
make the change of variables $x = \sqrt{N-1}(2r-1)$ and $y =
\sqrt{N-1}(2r'-1)$. This yields
\begin{equation}
\label{mvbdiu}
I_{0} \geq \sqrt{\frac{N-1}{N}} I(\alpha,\sqrt{N-1}) , 
\end{equation}
where 
\begin{eqnarray}
\label{defIab}
I(\alpha,\beta) = 
\int_{-\beta}^{\beta} \frac{e^{-\alpha \Phi (x)}}{1+x^{2}} 
\int_{-\beta}^{x}  
\frac{(-y)e^{\alpha \Phi (y)}}{(1+y^{2})^{3/2}} dydx   
\end{eqnarray}
with 
\begin{eqnarray}
\label{defalphalokal}
\alpha & = & \frac{T}{2LN^{\sigma-1}\sqrt{N-1}} = 
\frac{T}{2N^{\sigma}\arctan(\sqrt{N-1})} , 
\nonumber \\  
\Phi (x) & = & \int_{0}^{x}(1+x^{'2})^{\sigma-1}dx'.
\end{eqnarray} 
Furthermore 
\begin{eqnarray}
\label{dIdb}
\frac{d}{d\beta}I(\alpha,\beta) & = &   
- 2\frac{e^{-\alpha \Phi(\beta)}}{1+\beta^{2}}\int_{0}^{\beta}
\frac{y\sinh\biglb(\alpha \Phi(y)\bigrb)
}{(1+y^{2})^{3/2}}dy 
\nonumber \\ 
 & & + 2\frac{\beta e^{-\alpha \Phi(\beta)}
}{(1+\beta^{2})^{3/2}}\int_{0}^{\beta}\frac{\cosh\biglb(\alpha
\Phi(x)\bigrb)}{1+x^{2}}dx 
\nonumber \\ 
 & \geq & 2\frac{\beta e^{-\alpha \Phi(\beta)}}
{(1+\beta^{2})^{3/2}}\int_{0}^{\beta} 
\frac{e^{-\alpha \Phi(x)}}{1+x^{2}}dx 
\nonumber \\ 
 & \equiv & F(\alpha,\beta) > 0 . 
\end{eqnarray}
This expression is obtained by separating the integrals
``$\int_{-\beta}^{\beta}$'' into ``$\int_{-\beta}^{0}
+\int_{0}^{\beta} $'' and making the change of variables $x\rightarrow
-x$ and $y\rightarrow -y$ in the ``$\int_{-\beta}^{0}$'' integrals, as
well as by using the inequality
\begin{equation}
\label{ineqspec}
\frac{y}{(1+y^{2})^{(3/2)}}\leq
 \frac{\beta}{\sqrt{1+\beta^{2}}}\frac{1}{1+y^{2}}, \quad \forall y\in
 [0,\beta].
\end{equation}
It follows from Eq.~(\ref{dIdb}) that $I(\alpha,\beta)$ is increasing
in $\beta$, which together with Eqs.~(\ref{mvbdiu}) and (\ref{dIdb}) gives 
\begin{eqnarray}
I_{0} &\geq &\frac{1}{\sqrt{2}}I(\alpha,\sqrt{N-1}) \geq
\frac{1}{\sqrt{2}}I(\alpha,1)\nonumber\\ &\geq&
\frac{1}{\sqrt{2}}\int_{0}^{1} F(\alpha,\beta')d\beta' > 0,
\end{eqnarray}
where we have assumed that $N \geq 2$.
Thus, if $I_{0}\rightarrow 0$ then $\int_{0}^{1} F(\alpha,\beta') d\beta' 
\rightarrow 0$ necessarily. Furthermore, we have 
\begin{eqnarray}
 & & \frac{d}{d\alpha}\int_{0}^{1}F(\alpha,\beta)d\beta 
\nonumber \\ 
& = & -2 \int_{0}^{1} \left[ \frac{\beta \Phi (\beta) 
e^{-\alpha \Phi (\beta)}}{(1+\beta^{2})^{3/2}}\int_{0}^{\beta} 
\frac{e^{-\alpha \Phi (x)}}{1+x^{2}}dx \right.  
\nonumber\\
 & & \left. + \frac{\beta e^{-\alpha \Phi (\beta)}}
{(1+\beta^{2})^{3/2}}\int_{0}^{\beta} \frac{\Phi (x) 
e^{-\alpha \Phi (x)}}{1+x^{2}}dx \right] d\beta < 0. 
\nonumber \\  
\end{eqnarray}
It follows that $\int_{0}^{1} F(\alpha,\beta)d\beta$
is a strictly decreasing function in $\alpha$. Hence, a necessary
condition for this expression to go to zero is that $\alpha
\rightarrow
\infty$. For large $N$ it follows from the expression for $\alpha$ in 
Eq.~(\ref{defalphalokal}) that it is necessary for adiabaticity that
$T\gg N^{\sigma}$.

\begin{figure}
\includegraphics[width = 8.5cm]{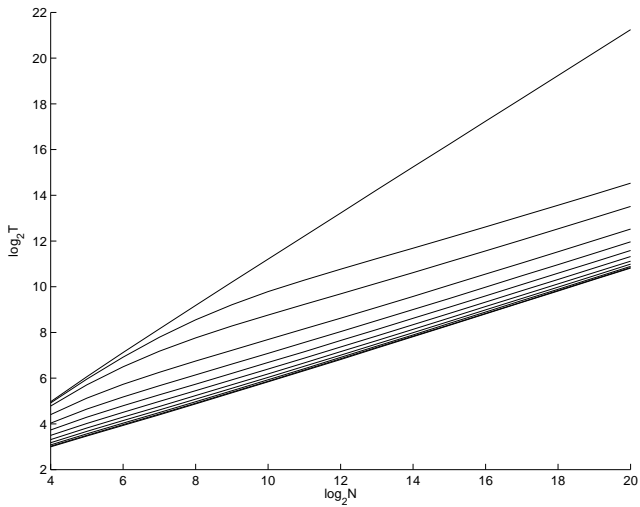}
\caption{\label{fig:lokalwfamilj} Local search with $W(r) = H(r)$ 
and success probability $0.5$. The curves show $\log_{2}T$
vs $\log_{2}N$, where $T$ is the run-time needed to obtain the
success probability $0.5$, and where $N$ is the list length. Each
curve shows the result for a given degree of decoherence
$\omega$. Counted from below, the curves correspond to $\omega =
0,0.1,\ldots,0.9,0.95,1$, interpolating between the closed ($\omega
=0$) and wide-open ($\omega=1$) case. As seen, all curves tend to the
slope $\frac{1}{2}$, except the uppermost wide-open case, which tends
to the slope $1$.}
\end{figure}

In Fig.~\ref{fig:lokalwfamilj}, we supplement the above analytic
results with numerical simulations of the dynamics of
Eq.~(\ref{eq:master}) with the choice $W(r)=H(r)$ and initial
condition $\rho_{00}(0) = 1$.  We interpolate between the closed and
the wide-open case, by letting $A=\cos (\omega\pi/2)$ and $B=\sin
(\omega\pi/2)$, where $\omega$ goes from $0$ to $1$. Furthermore, we
have assumed the success probability $\rho_{00}(1) = 0.5$. These
simulations confirm the predictions concerning the asymptotic
behavior, viz., that the evolution of the local adiabatic quantum
search stays near the instantaneous ground state if $T\gg
\sqrt{N}$ for all cases except the wide-open one, where $T\gg N$.

In conclusion, we have demonstrated that local adiabatic search is
robust to decoherence in the instantaneous eigenbasis of the search
Hamiltonian, as long as the Hamiltonian dynamics is present. Up to a
condition on the fluctuations, this result is independent of the
explicit form of the decoherence term.  This independence does no
longer hold in absence of the Hamiltonian part, in which case the
asymptotic behavior of the run-time of the local search changes. The
protective effect of the Hamiltonian dynamics is an indication of
robustness of quantum adiabatic search, which may be of importance in
physical implementations of a working search scheme that outperforms
any known classical search algorithm.

An interesting extension would be to apply the present analysis to
adiabatic algorithms designed to solve other problems, such as, e.g.,
the NP-complete problems 3-SAT \cite{farhi00} and exact cover
\cite{farhi01,Orus04,Latorre04}. Although analytical results may not
be achievable for these problems, numerical investigations could
reveal whether or not the protective effect of the Hamiltonian
dynamics is present. Moreover, one might consider whether there occurs
a transition from the seemingly polynomial behavior found in
\cite{farhi01}, to an exponential time-complexity, as the strength of
decoherence increases.

\vskip 0.1 cm We wish to thank Patrik Thunstr\"om for 
useful comments on the manuscript.


\begin{thebibliography}{99}
\bibitem{farhi00} E. Farhi, J. Goldstone, S. Gutmann, and M. Sipser, 
e-print quant-ph/0001106.
\bibitem{farhi01} E. Farhi, J. Goldstone, S. Gutmann, J. Lapan, 
A. Lundgren, and D. Preda, 
Science {\bf 292}, 472 (2001).
\bibitem{aharonov04} D. Aharonov, W. van Dam, J. Kempe, Z. Landau, S.
Lloyd, and O. Regev, e-print quant-ph/0405098; 
M. S. Siu, 
e-print quant-ph/0409024.
\bibitem{roland04} J. Roland and N. J. Cerf, 
e-print quant-ph/0409127.
\bibitem{kaminsky02} W. M. Kaminsky and S. Lloyd, 
e-print quant-ph/0211152.
\bibitem{childs01} A. M. Childs, E. Farhi, and J. Preskill, 
Phys. Rev. A {\bf 65}, 012322 (2001).
\bibitem{roland02} J. Roland and N. J. Cerf, 
Phys. Rev. A {\bf 65}, 042308 (2002).
\bibitem{vanDam01} W. van Dam, M. Mosca and U. Vazirani, in
{\it Proceedings of the 42nd Symposium on Foundations of Computer Science} 
(IEE Computer Society Press, New York, 2001), p.279.
\bibitem{paz99} Decoherence in the instantaneous energy eigenbasis 
is relevant, e.g., for a class of scenarios where the coupling 
to the environment is weak and dominated by the Hamiltonian 
of the system, see J. P. Paz and W. H. Zurek, 
Phys. Rev. Lett. {\bf 82}, 5181 (1999).
\bibitem{percival94} I. C. Percival, 
J. Phys. A {\bf 27}, 1003 (1994).
\bibitem{zanardi99} P. Zanardi and M. Rasetti, 
Phys. Lett. A {\bf 264}, 94 (1999).
\bibitem{Sarandy} M. S. Sarandy and D. A. Lidar,
Phys. Rev. A {\bf 71}, 012331 (2005).
\bibitem{Sarandy2} M. S. Sarandy and D. A. Lidar,
e-print quant-ph/0502014.
\bibitem{grover97} L. K. Grover, 
Phys. Rev. Lett. {\bf 79}, 325 (1997).
\bibitem{thecondition} As is evident from Eq.~(\ref{soloc}), improvement over
the classical time-complexities is obtained even for a weak $N$ dependence
in $K$, such as $K \propto N^a$, $a<\frac{1}{2}$.
\bibitem{aberg04} J. {\AA}berg, D. Kult, and E. Sj\"oqvist, 
(unpublished). 
\bibitem{sigmaremark}
The case where $\sigma < 1$ requires a different analysis. 
\bibitem{Orus04} R. Or\'{u}s and J. I. Latorre,
Phys. Rev. A {\bf 69}, 052308 (2004).
\bibitem{Latorre04} J. I. Latorre and R. Or\'{u}s, 
Phys. Rev. A {\bf 69}, 062302 (2004).







\end{thebibliography}
\end{document}